\documentclass[amsmath,aps,prb,reprint,superscriptaddress]{revtex4-1}

\usepackage{bbold}
\usepackage{graphics}
\usepackage{graphicx}
\usepackage{amsthm}
\usepackage{amsmath}
\usepackage{wasysym}
\usepackage{amssymb}
\usepackage{xfrac}
\usepackage{color}
\usepackage[dvips]{epsfig}
\renewcommand{\b}[1]{{\boldsymbol{#1}}}

\begin{document}

\bibliographystyle{naturemag}

\title{Fractionalized Topological Insulators}

\author{Joseph Maciejko}%
\email{maciejko@ualberta.ca}
\affiliation{Department of Physics, University of Alberta, Edmonton, Alberta T6G 2E1, Canada}
\affiliation{Theoretical Physics Institute, University of Alberta, Edmonton, Alberta T6G 2E1, Canada}
\affiliation{Canadian Institute for Advanced Research, Toronto, Ontario M5G 1Z8, Canada}
\author{Gregory A. Fiete}
\email{fiete@physics.utexas.edu}%
\affiliation{Department of Physics, The University of Texas at Austin, Austin, Texas 78712, USA}
\date{\today}

\begin{abstract}
Topological insulators have emerged as a major topic of condensed matter physics research with several novel applications proposed. Although there are now a number of established experimental examples of materials in this class,  all of them can be described by theories based on electronic band structure, which implies that they do not possess electronic correlations strong enough to fundamentally change this theoretical description.   Here, we review recent theoretical progress in the description of a class of strongly correlated topological insulators --- fractionalized topological insulators --- where band theory fails dramatically due to the fractionalization of the electron into other degrees of freedom.
\end{abstract}

\maketitle

The field of topological insulators (TI) is now approximately a decade old, and there are a number of excellent reviews available, from both an experimental and theoretical perspective\cite{Moore:nat10,Hasan:rmp10,Qi:rmp11,Ando:jpsj13}.  Our goal in this article is to provide a progress report on a specific subdirection that has developed rapidly in the last few years---the theoretical study of phases of matter that, while conceptually related to topological insulators, cannot be described by an electronic band structure because of the fractionalization of the electron into other degrees of freedom, driven by strong electronic correlations. This latter scenario has earlier precedents in the Mott insulator (which exhibits the separation of spin and charge) and the fractional quantum Hall effect (which has excitations with only a fraction of the charge of the electron), but acquires a distinct flavor when applied to topological insulators. By taking fractionalized topological insulators as an example, we hope to convey the broader message that the class of materials commonly known as topological insulators barely scratches the surface of the possible types of topological materials, once electron correlations are taken into account.

\section*{Splitting the electron apart}

We restrict our discussion to two and three spatial dimensions, as fractionalized excitations in gapped one-dimensional systems are subject to confining forces that prevent them from propagating over macroscopic distances. Since we focus on systems not described by band theory, we merely note in passing that interactions can drive a topologically trivial phase into a topological band insulator through the spontaneous generation of spin-orbit coupling \cite{Raghu:prl08,Zhang:prb09} or magnetic order\cite{yu2010}, or through the renormalization of electronic bands\cite{Zhang_Felser:sci}.  By contrast, in the phases we highlight below the quantum numbers of the electron break apart, implying that the latter is not a sharply defined quasiparticle and the system cannot be adequately described by an electronic band structure. The chief theoretical challenges for the study of fractionalized topological insulators are: (i) to describe them in a conceptually convenient way, (ii) to predict the conditions under which they are expected (i.e., find the set of Hamiltonians that support them), and if possible (iii) to provide a list of specific candidate materials that can be investigated experimentally. The field is presently at the stage of tackling challenges (i) and (ii), with relatively little work on (iii).   

A convenient way to describe the basic properties of fractionalized topological insulators is the slave-particle approach. In the present context, it answers the following question: What new state of matter does one get if electrons inside a topological insulator fractionalize into other degrees of freedom? The answer depends on the type of topological insulator with which one starts, as well as the assumed pattern of electron fractionalization. From these basic inputs, the slave-particle approach predicts the universal properties of the resulting fractionalized topological insulator, such as the quantum numbers and quantum statistics of low-energy excitations, the degeneracy of the ground state on surfaces of various topologies, and quantized electromagnetic, spin, and thermal response properties (Table~\ref{TableI}). The slave-particle approach does not, however, solve the much more difficult problem of the classification of all possible correlated topological phases---which goes beyond the scope of our focused review---because a generic correlated topological phase may not always be understood as originating via fractionalization from a noninteracting topological insulator. More sophisticated methods such as the group cohomology\cite{chen2012,mesaros2013} and cobordism\cite{kapustin2014} approaches have been recently developed to address this considerable challenge.

Besides fractionalization, to obtain an exotic topological phase one typically assumes that some of the fractionalized degrees of freedom form a topological band insulator. The latter can thus be seen as a nonfractionalized parent state from which fractionalized daughter states can be constructed (Table~\ref{TableI}, first two columns). The nontrivial response properties of the parent state combine with the fractional quantum numbers of the fractionalized degrees of freedom to yield even more exotic response properties for the daughter state. This scenario occurs in slave-particle mean-field studies of models of topological band insulators subjected to strong electron-electron interactions. Besides interacting with external fields via their fractional quantum numbers such as spin or charge, the fractionalized degrees of freedom also interact with emergent gauge fields. As in quantum chromodynamics (QCD), these gauge fields are confining in a conventional phase where fractionalized degrees of freedom---``quarks'' in the QCD analogy---do not exist at low energies as long-lived, propagating excitations. Inside a fractionalized topological insulator, the gauge fields are deconfined and allow the fractionalized degrees of freedom to lead an existence of their own. The gauge fields themselves admit propagating excitations---the gauge bosons---which can be gapless or gapped, depending on the type of gauge field and the nature of the parent topological phase.

\begin{widetext}

\begin{table}
\begin{tabular}{|c|c|c||c|c|c|c|c|}
\hline
parent state & fractionalized state &  $d$ & symmetry & EM response & spin response & thermal response \\
\hline
CI & CSL & 2 &  & $\sigma_{xy}=0$ & $\sigma_{xy}^s=0$ & $\kappa_{xy}/T=2$  \\
QSH & fractionalized QSH & 2 & TRS & $\sigma_{xy}=0$ & $\sigma_{xy}^s=0$ & $\kappa_{xy}/T=0$ \\
TI/WTI & TMI/WTMI & 3 & TRS & $\theta=0$ & & $\theta_\textrm{grav}=\pi$ (TMI) \\
TCI & TCMI & 3 & mirror &  & & \\
\hline
CI & CI* & 2 &  & $\sigma_{xy}=2$ & $\sigma_{xy}^s=0$ & $\kappa_{xy}/T=2$ \\
QSH & QSH* & 2 & TRS & $\sigma_{xy}=0$ & $\sigma_{xy}^s=2$ & $\kappa_{xy}/T=0$ \\
TI & TI* & 3 & TRS & $\theta=\pi$ & & $\theta_\textrm{grav}=\pi$ \\
\hline
CI & FCI & 2 &  & $\sigma_{xy}=1/3$ & $\sigma_{xy}^s=0$ & $\kappa_{xy}/T=1$\\
QSH & FQSH & 2 & TRS & $\sigma_{xy}=0$ & $\sigma_{xy}^s=2/3$ & $\kappa_{xy}/T=0$\\
TI & FTI & 3 & TRS & $\theta=\pi/3$ & & $\theta_\textrm{grav}=\pi$\\
\hline
\end{tabular}
\caption{Summary of the $d$-dimensional fractionalized topological insulators discussed in this article and their universal response properties (acronyms are defined in the main text). $\sigma_{xy}$: Hall conductance in units of $e^2/h$; $\theta$: coefficient of the $\b{E}\cdot\b{B}$ term in units of $e^2/2\pi h$; $\sigma_{xy}^s$: spin Hall conductance in units of $e/4\pi$ for a system with conserved $z$ component of spin; $\kappa_{xy}/T$: thermal Hall conductance divided by the temperature, in units of $\pi^2 k_B^2/3h$; $\theta_\textrm{grav}$: coefficient of the gravitational equivalent of the $\b{E}\cdot\b{B}$ term, responsible for a surface quantized thermal Hall effect.}
\label{TableI}
\end{table}

\end{widetext}

\section*{Fractionalization and new phases of matter}

\begin{figure}
\includegraphics[width=1\linewidth]{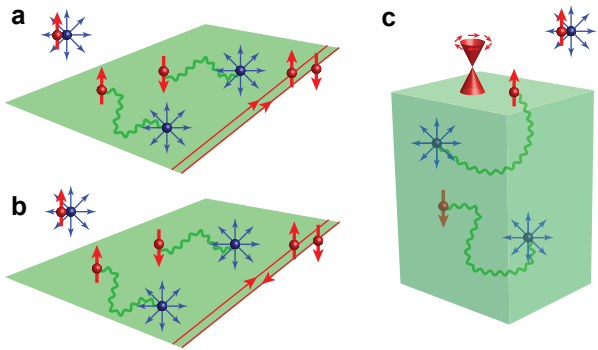}
\caption{Fractionalized topological insulators accessed by the $U(1)$ slave-rotor description. Outside the material, the charged bosonic rotor (blue sphere) and the neutral fermionic spinon (red sphere) are confined together to form an electron. Inside the material, they become deconfined particles and interact via an emergent $U(1)$ gauge field (green wiggly line). In the 2D chiral spin liquid (a), these particles acquire semionic (i.e., half-fermionic) statistics and there are chiral edge states of spinons. In the 2D fractionalized quantum spin Hall insulator (b), the spinon edge states are helical. In the 3D topological Mott insulator states (c), the $U(1)$ gauge field is gapless in the bulk, and there are gapless Dirac-like spinon surface states.}
\label{fig:fig1}
\end{figure}

The first pattern of electron fractionalization to be considered in the context of strongly correlated topological insulators was spin-charge separation, originated in the study of the Luttinger liquid in 1D and the Mott transition in 2D\cite{florens2004,Lee:rmp06}. The electron is assumed to fractionalize into a charged, spinless, bosonic rotor and an electrically neutral, spinful, fermionic spinon, coupled by a $U(1)$ gauge field (Fig.~\ref{fig:fig1}). The electronic quasiparticle weight is proportional to the expectation value of the rotor field. The resulting theory typically has two distinct stable phases: a weakly interacting phase where the rotors undergo Bose condensation and the rotor field acquires a nonzero expectation value, and a strongly interacting phase where the rotors are uncondensed. In the weakly interacting phase, the electronic quasiparticle weight is finite and the electronic band structure is well defined, while in the strongly interacting phase the quasiparticle weight vanishes due to strong rotor fluctuations, corresponding to a Mott phase. In the latter case, the spinons acquire a band structure of their own which, if gapped, can be classified by the usual topological invariants\cite{Hasan:rmp10,Qi:rmp11}. In 2D, Chern-insulator (CI) or $\mathbb{Z}_2$ topological insulator band structures for the spinons lead to a chiral spin liquid\cite{he2011} (CSL) or a fractionalized quantum spin Hall (QSH) insulator\cite{young2008,rachel2010}, respectively. The fractionalized QSH insulator is only stable in the presence of additional layers containing enough gapless spinons to screen the $U(1)$ gauge field and avoid the instanton effect that causes confinement\cite{young2008}. In 3D time-reversal invariant systems, the Mott phases are called strong topological Mott insulators\cite{Pesin:np10} (TMI) or weak topological Mott insulators\cite{Kargarian:prb11} (WTMI) depending on whether the spinons form a strong or weak $\mathbb{Z}_2$ topological insulator, respectively. The same analysis applied to systems with mirror crystalline symmetries leads to the concept of topological crystalline Mott insulator\cite{Kargarian:prl13} (TCMI). All these 3D fractionalized topological phases have gapless surface modes of spinons with the properties that electrons would have in the corresponding noninteracting topological phases, but in addition have a bulk gapless ``photon" mode from the emergent $U(1)$ gauge field\cite{Pesin:np10,Witczak-Krempa:prb10}. However, there is no quantized electromagnetic response of the $\b{E}\cdot\b{B}$ type\cite{Moore:nat10,Hasan:rmp10,Qi:rmp11} because the spinons are electrically neutral. To identify this class of spin-charge separated states experimentally, methods similar to those proposed or used to identify quantum spin liquids\cite{Balents:nat10} should be applied. One should observe a bulk contribution to the low-temperature specific heat arising from the gapless photon mode, although it could be difficult to distinguish from the phonon contribution. In addition, one could imagine detecting the linearly dispersing surface spinon modes via the observation of 2D spin or heat transport localized at the surface of the material.

\begin{figure}
\includegraphics[width=1\linewidth]{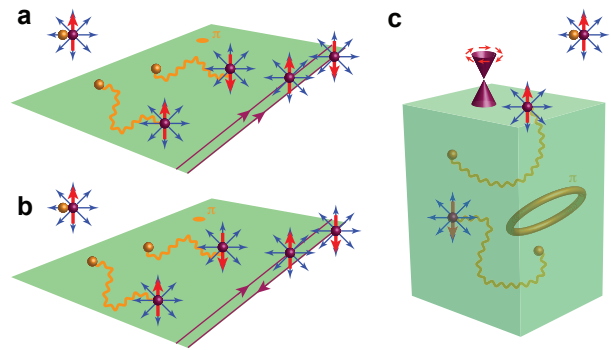}
\caption{Fractionalized topological insulators accessed by the $\mathbb{Z}_2$ slave-Ising description. Outside the material, the neutral bosonic slave-Ising particle (orange sphere) and the charged slave-fermion (purple sphere) are confined together to form an electron. Inside the material, they become deconfined particles and interact via an emergent $\mathbb{Z}_2$ gauge field (orange wiggly line) that supports $\pi$-flux vortex excitations (orange disk). In the 2D correlated Chern insulator (a), vortices acquire semionic statistics and there are chiral edge states of slave-fermions. In the 2D correlated quantum spin Hall insulator (b), vortices have semionic mutual statistics with slave-Ising particles and slave-fermions, and the slave-fermion edge states are helical. In the 3D correlated topological insulator (c), the $\pi$-flux vortex becomes a bulk vortex loop, and there are gapless Dirac-like slave-fermion surface states.}
\label{fig:fig2}
\end{figure}

A different pattern of electron fractionalization originated from theories of high-temperature superconductivity\cite{senthil2000} assumes that the electron fractionalizes into a neutral, spinless, bosonic slave-Ising particle and a charged, spinful slave-fermion, coupled by a $\mathbb{Z}_2$ gauge field\cite{ruegg2010} (Fig.~\ref{fig:fig2}). This type of fractionalization may appear trivial, as the slave-Ising particle carries neither the charge nor the spin of the electron. The resulting fractionalized phases, however, are far from trivial. As in the $U(1)$ theory, the $\mathbb{Z}_2$ theory typically has a non-fractionalized weakly interacting phase where the slave-Ising particles Bose condense, and a fractionalized strongly interacting phase where they are uncondensed. By contrast with the $U(1)$ theory however, in the $\mathbb{Z}_2$ case the slave-fermion carries the charge of the electron, thus the fractionalized phases inherit the electromagnetic response properties of the parent topological band insulator. In 2D, Chern-insulator and $\mathbb{Z}_2$ topological insulator band structures for the slave fermions lead to a correlated Chern insulator\cite{Maciejko:prb13} (CI*) with quantized Hall response and a correlated quantum spin Hall insulator\cite{Ruegg:prl12} (QSH*), respectively. These are fully gapped phases with intrinsic topological order of the double-semion and toric-code type, respectively. In 3D time-reversal invariant systems, a $\mathbb{Z}_2$ topological insulator band structure leads to a strongly correlated topological insulator\cite{Maciejko:prl14} (TI*) that, unlike the TMI described above, exhibits a quantized $\b{E}\cdot\b{B}$ electromagnetic response and has no gapless bulk excitations. The TI* is a true 3D topological phase whose bosonic sector is described by a topological $BF$ theory\cite{hansson2004,Cho:ap10} that imparts a relative statistical angle of $\pi$ between bulk particle and loop excitations and is responsible for a nontrivial topological ground-state degeneracy.

\begin{figure}
\includegraphics[width=1\linewidth]{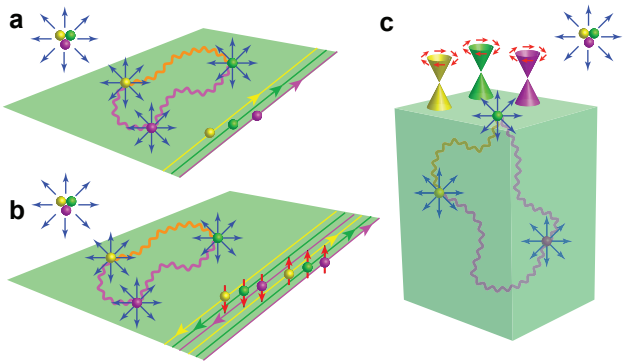}
\caption{Fractional topological insulators. Outside the material, the three fermionic partons of charge $e/3$ are confined together to form an electron. Inside the material and in the $U(1)\times U(1)$ description, they become deconfined particles and interact via two $U(1)$ gauge fields (orange and magenta wiggly lines). In the 2D phases (a) and (b), the partons acquire fractional statistics. In the fractional Chern insulator (a), there are chiral edge states of partons, while in the fractional quantum spin Hall insulator (b) the parton edge states are helical. In the 3D fractional topological insulator (c), the $U(1)$ gauge fields are gapless in the bulk, and there are gapless Dirac-like parton surface states. In the $\mathbb{Z}_3$ description, the gauge field would be gapped in the bulk with vortex loop excitations similar to those of Fig.~\ref{fig:fig2}(c).}
\label{fig:fig3}
\end{figure}

Finally, there are even more exotic fractionalized phases in which the electron with charge $e$ splits apart into three fermionic partons of charge $e/3$ (Fig.~\ref{fig:fig3}). In 2D, when these partons form noninteracting integer quantum Hall states one obtains the familiar fractional quantum Hall states. In the conceptually similar case of a Chern insulating band structure, one obtains fractional Chern insulators (FCI)---fractional quantum Hall states that exist in the absence of an external magnetic field. There is already a sizeable literature on FCI, and we direct the reader to existing review articles on the subject\cite{parameswaran2013,bergholtz2013} for more information and original references. If the fractionally charged fermions form a time-reversal invariant $\mathbb{Z}_2$ topological insulator, one obtains a fractional quantum spin Hall insulator or 2D fractional topological insulator\cite{Bernevig:prl06,Levin:prl09,karch2010,lu2012,chan2013} (FTI), and in 3D a 3D FTI\cite{Maciejko:prl10,Swingle:prb11,hoyos2010,
Maciejko:prb12}. The relevant gauge theories in the example above are of the $U(1)\times U(1)$ or $\mathbb{Z}_3$ variety, leading to gapless bulk photons or no bulk gapless excitations, respectively (a non-Abelian $SU(3)$ gauge theory is formally possible but is typically plagued by the problem of confinement). In both cases the 3D FTI exhibits a quantized $\b{E}\cdot\b{B}$ electromagnetic response, but with a coefficient one-third that of a topological band insulator. From an experimental standpoint, a 3D FTI should be easier to detect than a 3D TMI because its gapless surface states are electrically charged, leading to nontrivial signatures in photoemission, tunneling, and quantum oscillations\cite{Swingle:prb12}. 

Can the fractionalized topological insulators whose phenomenology we have described so far be reliably obtained as ground states of interacting electronic Hamiltonians? The answer is a partial yes. In 2D, FCI ground states have been reliably obtained as ground states of models of spinless electrons with nearest-neighbor repulsive interactions via numerical studies\cite{parameswaran2013,bergholtz2013} and strong-coupling expansions\cite{mcgreevy2012}. 2D time-reversal invariant FTI ground states have likewise been found in numerical studies\cite{neupert2011b,repellin2014} and exactly soluble models\cite{levin2011,koch-janusz2013} of interacting fermions and/or bosons (compressible, bosonic analogs of 2D FTI may have also been found in microscopic models of interacting bosons\cite{motrunich2007}). Progress has also been made in systematically designing Hamiltonians for which various spin-charge separated states such as the CSL are exact ground states\cite{schroeter2007,mei2014}, but these are models of interacting localized spins rather than models of interacting electrons. The challenge remains mostly open in 3D---where numerical studies come at a prohibitive computational cost---with the exception of exactly soluble models\cite{levin2011}.

Do the exotic types of order found in fractionalized topological insulators compete or coexist with conventional symmetry-breaking orders such as magnetism, superconductivity, and charge-density waves? On the one hand, if the fractionalized phase is protected by a symmetry that would be broken by the onset of conventional order, then coexistence is ruled out. For example, time-reversal invariant FTI would not generally coexist with magnetic order---unless magnetic order occurs only on the edge, which would preserve fractionalization in the bulk. On the other hand, if conventional order does not break the protecting symmetry, then coexistence is generally allowed, but need not occur.

We have discussed a few examples of fractionalized topological insulators. While a list of specific materials that would harbor such phases is currently lacking, materials that combine the fundamental ingredients of strong spin-orbit coupling and strong electron correlations---such as the pyrochlore iridates\cite{witczak-krempa2014} and samarium hexaboride\cite{dzero2013}---do exist and are currently under intense experimental study. Such materials, as well as systems of ultracold atoms subjected to synthetic gauge fields, may unveil entirely novel quantum phases of matter, topological or otherwise.

Correspondence and requests for materials should be addressed to J.M. We are grateful to our collaborators in this area, V. Chua, A. Karch, M. Kargarian, X.-L. Qi, A. R\"uegg, T. Takayanagi, C.-C. J. Wang, J. Wen, and S.-C. Zhang. Our work was generously funded by the ARO, DARPA, DOE, NSF, the Simons Foundation, NSERC, CRC, CIFAR, and startup funds from the University of Alberta. J.M. and G.A.F. contributed equally to this work.

\end{document}